\documentclass[runningheads,citeauthoryear]{apinv_kurtedit}
\usepackage{epsfig,cite,graphics}
\usepackage{marvosym} 
\usepackage[utf8]{inputenc}
\usepackage{amssymb,amsmath}

\begin{document}

\title{Ultra-short period contact binaries: restricting the parameters of the primary using Gaia parallax}
\titlerunning{Ultra-short period contact binaries: restricting the parameters of the primary}
\author{Alexander Kurtenkov\inst{1}}
\authorrunning{A. Kurtenkov}
\tocauthor{Alexander Kurtenkov} 
\institute{Institute of Astronomy and NAO, Bulgarian Academy of Sciences, \\ 72 Tsarigradsko Shose Blvd., BG-1784, Sofia, Bulgaria   \newline
	\email{al.kurtenkov@gmail.com}    }
\papertype{Submitted on 20.08.2021; Accepted on 02.12.2021}	
\maketitle

\begin{abstract}
Possible M-type contact binaries were investigated by selecting W UMa-type variables with orbital periods below the 0.22-day cutoff. Gaia parallaxes were combined with Gaia and 2MASS photometry to obtain G and J-band absolute magnitudes of 674 red variable stars catalogued as contact binaries in the VSX database. The absolute magnitude of main sequence cool dwarfs varies strongly with spectral type, which was used to create a selection of 218 systems with primaries potentially of spectral types M0-M3. Lightcurves of the 46 systems with lowest near-infrared luminosities were inspected individually to confirm or reject the W UMa classification where possible. The extinction limits and amplitudes were combined with the calculated absolute magnitudes in order to set upper limits for the masses of the primary components. This is achieved via the consideration that the luminosity of the primary is no less than that of a main sequence star of the same mass. For 26 possible contact binaries the mass of the primary was limited to less than 0.5 solar masses with some systems having upper limits for the primary as low as 0.35 solar masses. These rare systems are intriguing targets for more detailed observations since they are on the low-mass end of the contact binary distribution and their formation and evolution are still unclear.    
\end{abstract}

\section*{1. Introduction and rationale}
Contact binary systems are predominantly of spectral types F, G, and K. They are usually observed as variable stars of the W UMa-type: eclipsing variables whose lightcurves present round-shaped maxima as well as primary and secondary minima of similar depths due to the small temperature difference of the components. Notable models explaining the W UMa lightcurves as caused by eclipses in contact binary systems include those by Lucy (1968) and Rucinski (1993). Two main W UMa subtypes are defined (Binnendijk 1970) -- the A-subtype, usually consisting of A and F spectral type components, and the W-subtype, which have G and K type components and, respectively, shorter orbital periods. Only a few confirmed M-type contact binaries have been discovered and even fewer of these are well-studied (Latkovi{\'c} et al. 2021). The period distribution of contact binaries has a sharp cut-off at $P=0.22$\,d (Rucinski 1992). As M-dwarfs have the lowest radii on the stellar main sequence, M-type systems are expected to have the lowest semi-major axes and, respectively, the shortest periods among contact binaries. Due to the rarity of these M-type systems, variables, classified as W UMa-type binaries below the 0.22-day period cut-off are of great interest. It is still an open question why these M-type binaries are so rare even after accounting for the selection bias caused by their low luminosity. Jiang et al. (2012) suggest that systems with a primary component mass below 0.63\,\msol\, are dynamically unstable upon activation of the mass transfer through $L_{1}$, which hinders the formation of a durable low-mass contact binary. This consideration elevates the interest specifically in systems with low-mass primaries.

The current work aims to select possible M-type contact binaries with low-mass primaries for further, more detailed studies. The number of known contact binaries has considerably increased thanks to recent variability surveys, most importantly, the Zwicky Transient Facility (Chen et al. 2020). Meanwhile, the Gaia mission (Gaia Collaboration et al. 2021) has supplied parallax data for systems at a distance of the order of 1\,kpc with an unprecedented reliability. As M-type discoveries are expected at low distances due to their low luminosities, Gaia data combined with other photometry can be used to calculate approximate absolute magnitudes. The absolute magnitudes vary strongly for M-dwarfs -- the typical difference between M0V and M5V is 3.1\,mag in J-band and 4.3\,mag in Gaia G-band (Pecaut \& Mamajek 2013). Thus, these absolute magnitudes can be used to set an upper limit of the luminosity and spectral type of the primary -- the extremal case being if the luminosity of the secondary is negligible. As there is no reason for the primary to be less luminous than a single main sequence star of the same mass, an upper limit of the mass can also be estimated. It is hereby suggested that this is a decent method to restrict the parameters of the M-type primary, whereas colors are more strongly influenced by the uncertainties due to the amplitude of the eclipses and the interstellar extinction relative to their differences between spectral subtypes.

\section*{2. Object selection}
As of 2021 the Variable Star Index (VSX\footnote{\url{https://www.aavso.org/vsx/}}) lists 2489 objects, classified as W UMa-type variables. The following criteria were used when matching these to the Gaia EDR3 catalog:
\begin{itemize}
    \item Relative Gaia parallax error $e\_Plx/Plx < 0.2$
    \item Gaia $RP$ magnitude error $e\_RPmag < 0.1$
    \item Gaia color $G-RP > 0.8$
\end{itemize}

The latter value has been chosen as the typical value for type M0V is $G-RP=0.9$, whereas the other criteria are liberal. A total of 688 matches were found. These were matched with the 2MASS catalog in order to include the $J$-band magnitudes. The only criterion was the error in $J$-band magnitude $e\_Jmag<0.2$. This left us with 674 objects. A search radius of $5''$ was used for all matches.

Initial absolute magnitudes (uncorrected for extinction) were calculated by applying the distance moduli calculated from the Gaia parallax $p$ via:
\begin{eqnarray}
    M_{G}=G+5\lg(p\textrm{[mas]})-10                            \\
    M_{J}=J+5\lg(p\textrm{[mas]})-10
\end{eqnarray}

Although it would be difficult to determine the actual extinction, the maximum values can be set at the galactic line-of-sight extinction as calculated by Schlafly \& Finkbeiner (2011). As these are maximum values, they are not applied to the absolute magnitudes, but they are presented next to the results for consideration. A major advantage of using the 2MASS $J$-band magnitudes is the low extinction in the near infrared. 

\begin{table}[htb]
  \begin{center}
  \caption{Typical parameters of main sequence M-dwarfs.}
  \begin{tabular}{lcccccc}
  \hline
  Spectral type \,\, & $T_{eff}$ \,\, & $R[R_{\sun}]$ \,\, & Mass $[M_{\sun}]$ \,\, & $BP-RP$ \,\, & $M_G$ \,\, & $M_J$ \,\, \\
  \hline
  M0V   & 3850 & 0.588 & 0.57 & 1.84 & 8.16 \,& 5.97 \\
  M0.5V & 3770 & 0.544 & 0.54 & 1.97 & 8.44 \,& 6.19 \\
  M1V   & 3660 & 0.501 & 0.50 & 2.09 & 8.82 \,& 6.48 \\
  M1.5V & 3620 & 0.482 & 0.47 & 2.13 & 8.98 \,& 6.59 \\
  M2V   & 3560 & 0.446 & 0.44 & 2.23 & 9.29 \,& 6.81 \\
  M2.5V & 3470 & 0.421 & 0.40 & 2.39 & 9.67 \,& 7.01 \\
  M3V   & 3430 & 0.361 & 0.37 & 2.50 & 10.05 \,& 7.38 \\
  M3.5V & 3270 & 0.300 & 0.27 & 2.78 & 10.87 \,& 7.93 \\
  M4V   & 3210 & 0.274 & 0.23 & 2.94 & 11.21 \,& 8.20 \\
  M4.5V & 3110 & 0.217 & 0.18 & 3.16 & 12.04 \,& 8.80 \\
  M5V   & 3060 & 0.196 & 0.16 & 3.35 & 12.45 \,& 9.09 \\
  M5.5V & 2930 & 0.156 & 0.12 & 3.71 & 13.35 \,& 9.72 \\
  M6V   & 2810 & 0.137 & 0.10 & 4.16 & 14.26 \,& 10.18 \\
  M6.5V & 2740 & 0.126 & 0.09 & 4.50 & 14.40 \,& 10.47 \\
  M7V   & 2680 & 0.120 & 0.09 & 4.65 & 14.72 \,& 10.70 \\
  \hline
  \end{tabular}
  \label{Mamajek}
  \end{center}
\end{table} 

Typical parameters of main sequence M-dwarfs after Pecaut \& Mamajek (2013), supplemented online\footnote{\url{https://www.pas.rochester.edu/\~emamajek/EEM\_dwarf\_UBVIJHK\_colors\_Teff.txt}} were used to set a limit for the spectral type of the primary component. The values are presented in Table\,\ref{Mamajek}. A list of possible M-type contact binaries was selected based on these values by applying the following criteria:
\begin{itemize}
    \item Absolute $G$-band magnitude $M_G > 8.2$
    \item Absolute $J$-band magnitude $M_J > 6.0$
    \item Gaia color $BP-RP > 1.9$
\end{itemize}

\begin{figure}[!ht]
\centering
\includegraphics[width=6.2cm]{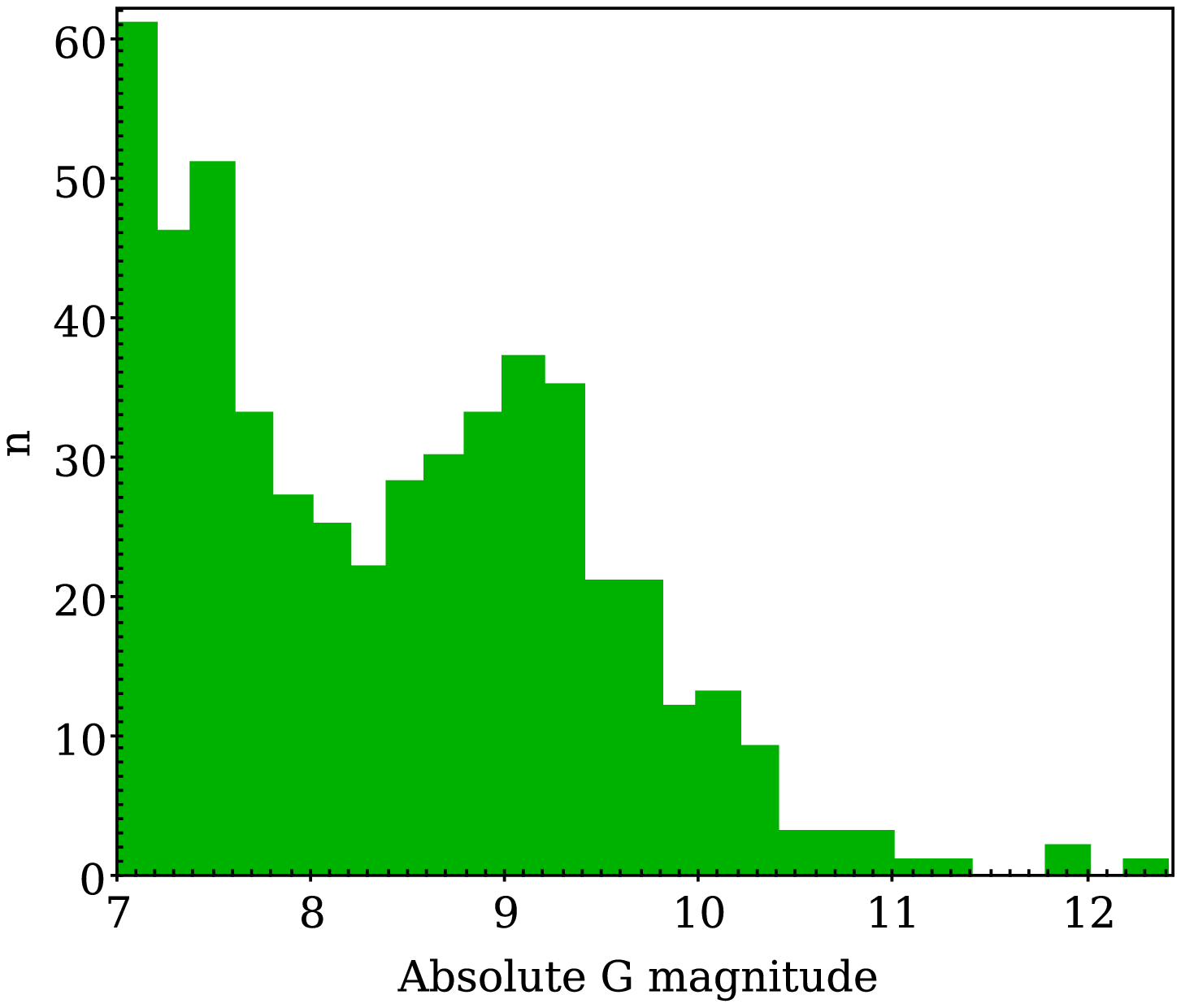}
\includegraphics[width=6.2cm]{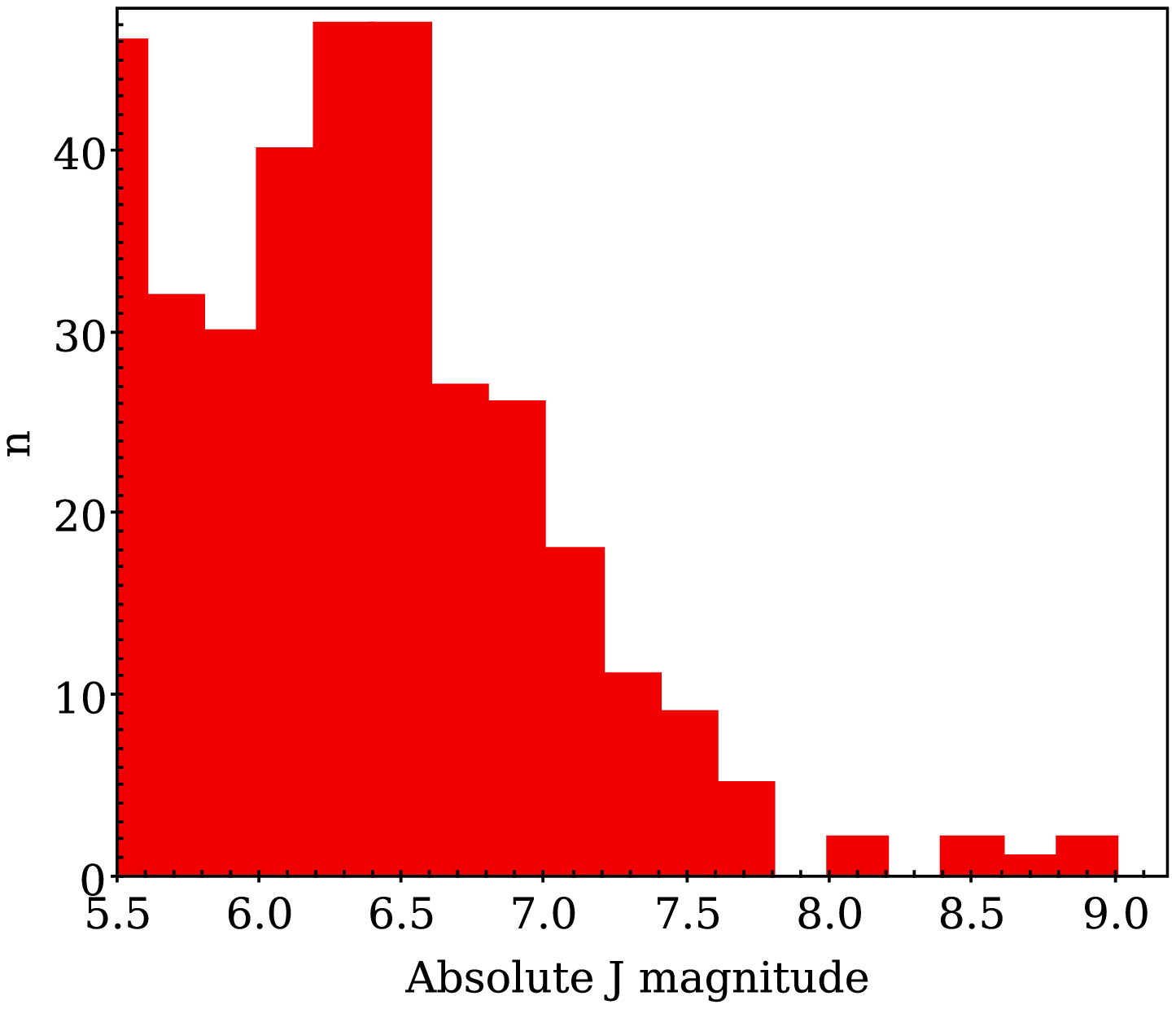}
\caption{Absolute magnitude distribution in $G$ and $J$-band using 0.2\,mag bins. Not corrected for extinction.}
\label{fig_distrib}
\end{figure}

This selection contains 218 systems, which are of significant interest for further observations and analysis, e.g. lightcurve modeling. The majority of these systems, if they have been classified correctly, are expected to contain a primary less massive than $0.6M_{\sun}$. The cutoff values of the absolute magnitude distributions ($\approx7.5$ for $M_{J}$ and $\approx10$ for $M_{G}$, Fig.\,\ref{fig_distrib}) suggest that almost all of these systems are of spectral type earlier than M3.5V, so masses of the primaries below $\approx0.3M_{\sun}$ should be extremely rare if there are any.  

\begin{figure}[!ht]
\centering
\includegraphics[width=12.0cm]{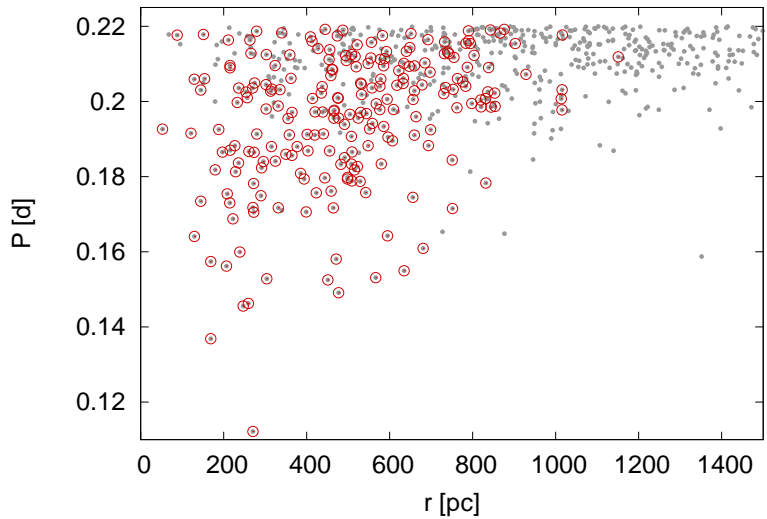}
\caption{A distance-period diagram for all 674 systems (small grey circles). The selected 218 systems (large hollow red circles) are statistically closer to the Earth as their luminosities are lower and they include most of the systems with the shortest periods.}
\label{fig_period}
\end{figure}

The orbital periods of the selected systems are plotted versus distance on Fig.\,\ref{fig_period}. It shows that the selected systems have statistically shorter periods, as their radii are smaller. Due to their low luminosities almost all of them are within 1\,kpc from the Earth, whereas many systems from the initial selection are located farther.

The full table containing coordinates, lightcurve parameters and magnitudes of the catalog matches for all systems is available upon request from the author.

\section*{3. Objects near the low-mass end}

The selection contains 46 objects with $M_{J}>7.0$, which should be near the low-mass end. These objects are presented in Table\,\ref{lowmass1} with their amplitudes $A$ and periods $P$ according to VSX and their Gaia parallaxes $p$ and $BP-RP$ colors. Each of them was individually considered as an object of interest. 

\begin{table}[!htb]
\scriptsize
  \begin{center}
  \caption{The list of selected M-type contact binary candidates with the lowest $J$-band luminosity: parameters from VSX and Gaia.}
  \begin{tabular}{lccccccc}
  \hline
   Name                \, &      RA (J2000)  \, &    Dec (J2000)  \, &    $A$\,[mag]  \, & $P$\,[d]  \, &  $p$\,[mas] \, &  $\sigma_{p}$  \, &  $BP-RP$  \\           
  \hline
   ZTF J024005.10+513845.5 \, & 40.0212548   &  51.64594  &   0.409 \, &   0.1599  \, &   4.187   &   0.835  &   2.57 \\
   WTS 19b-3-06008         \, & 293.455638   &  36.90054  &   0.08  \, &   0.1121  \, &   3.697   &   0.493  &   2.57 \\
  CSS J093044.8+264957     \, & 142.686717   &  26.83252  &   0.2   \, &   0.1970  \, &   2.742   &   0.484  &   2.01 \\
   ZTF J150926.26+583628.1 \, & 227.359485   &  58.60779  &   0.161 \, &   0.1734  \, &   6.911   &   0.057  &   2.78 \\
   ZTF J001012.42+452925.1 \, & 2.55179707   &  45.49026  &   0.179 \, &   0.1836  \, &   4.243   &   0.116  &   2.57 \\
   ZTF J213409.46+335033.8 \, & 323.539439   &  33.84273  &   0.534 \, &   0.2095  \, &   4.644   &   0.346  &   2.45 \\
   ZTF J191844.32+532028.2 \, & 289.684684   &  53.34113  &   0.289 \, &   0.1717  \, &   3.016   &   0.085  &   2.40 \\
   ZTF J233629.58+505049.5 \, & 354.123284   &  50.84707  &   0.153 \, &   0.1818  \, &   5.570   &   0.060  &   3.08 \\
   ZTF J204919.36+080812.0 \, & 312.330697   &  8.136691  &   0.142 \, &   0.1910  \, &   2.791   &   0.265  &   2.13 \\
   ZTF J175927.14+654540.4 \, & 269.863068   &  65.76120  &   0.579 \, &   0.1527  \, &   3.292   &   0.109  &   2.76 \\
  CSS J234019.3+122438     \, & 355.080936   &  12.41074  &   0.35  \, &   0.1368  \, &   5.924   &   0.056  &   2.36 \\
   ZTF J174317.23+401313.1 \, & 265.821783   &  40.22024  &   0.718 \, &   0.1490  \, &   2.096   &   0.202  &   2.51 \\
   ZTF J145410.58+314216.2 \, & 223.544137   &  31.70448  &   0.232 \, &   0.2060  \, &   6.503   &   0.051  &   2.68 \\
   ZTF J193121.28+293951.2 \, & 292.838716   &  29.66423  &   0.293 \, &   0.1580  \, &   2.122   &   0.174  &   2.72 \\
   ZTF J051637.22+380939.8 \, & 79.1544408   &  38.16008  &   0.234 \, &   0.2043  \, &   1.475   &   0.256  &   1.97 \\
   ZTF J220421.03+163655.8 \, & 331.087656   &  16.61549  &   0.243 \, &   0.1748  \, &   3.452   &   0.130  &   2.99 \\
   ZTF J174214.33+642525.0 \, & 265.559742   &  64.42358  &   0.283 \, &   0.2025  \, &   3.941   &   0.062  &   2.75 \\
   ZTF J182843.57+372540.6 \, & 277.181579   &  37.42795  &   0.28  \, &   0.2131  \, &   1.351   &   0.174  &   2.21 \\
   ZTF J011509.92+130340.9 \, & 18.7913604   &  13.06135  &   0.462 \, &   0.1549  \, &   1.575   &   0.267  &   2.07 \\
   ZTF J015312.72+444309.8 \, & 28.3029965   &  44.71939  &   0.4   \, &   0.1754  \, &   4.793   &   0.099  &   2.87 \\
   ZTF J003651.36+503617.7 \, & 9.21400193   &  50.60490  &   0.414 \, &   0.1788  \, &   1.964   &   0.173  &   2.61 \\
   ZTF J175432.47+373447.6 \, & 268.635354   &  37.57988  &   0.221 \, &   0.1925  \, &   5.315   &   0.047  &   2.90 \\
   ZTF J053128.93+512221.6 \, & 82.8705820   &  51.37263  &   0.41  \, &   0.1959  \, &   1.507   &   0.261  &   2.25 \\
  CSS J082958.5+355923     \, & 127.494109   &  35.98983  &   0.22  \, &   0.1455  \, &   4.050   &   0.095  &   2.71 \\
   ZTF J011434.64+622754.1 \, & 18.6443964   &  62.46503  &   0.181 \, &   0.1705  \, &   3.671   &   0.067  &   2.65 \\
   ZTF J221742.28+485418.4 \, & 334.426184   &  48.90513  &   0.156 \, &   0.1954  \, &   2.816   &   0.088  &   2.67 \\
   ZTF J183104.30+712050.5 \, & 277.767912   &  71.34735  &   0.157 \, &   0.1879  \, &   3.172   &   0.065  &   2.39 \\
  CSS J111647.8+294602     \, & 169.199264   &  29.76744  &   0.3   \, &   0.1462  \, &   3.852   &   0.080  &   2.47 \\
   ZTF J001725.54+513930.3 \, & 4.35645188   &  51.65844  &   0.348 \, &   0.2008  \, &   2.101   &   0.156  &   2.69 \\
   2MASS J08115860+3119595 \, & 122.994093   &  31.33335  &   0.3   \, &   0.1561  \, &   4.839   &   0.050  &   2.26 \\
   ZTF J112409.81+314128.5 \, & 171.040919   &  31.69124  &   0.293 \, &   0.1913  \, &   2.271   &   0.192  &   2.62 \\
  CSS J224134.8+253648     \, & 340.395580   &  25.61351  &   0.29  \, &   0.1573  \, &   5.928   &   0.064  &   2.86 \\
  CSS J030238.1-065034     \, & 45.6587354   &  -6.84290  &   0.19  \, &   0.2128  \, &   3.766   &   0.086  &   2.87 \\
  CSS J001242.4+130809     \, & 3.17693974   &  13.13583  &   0.24  \, &   0.1640  \, &   7.730   &   0.047  &   2.87 \\
   ZTF J184257.77-014740.2 \, & 280.740741   &  -1.79452  &   0.422 \, &   0.1705  \, &   2.508   &   0.142  &   2.70 \\
   ZTF J191410.07+322037.4 \, & 288.542008   &  32.34370  &   0.208 \, &   0.2048  \, &   1.886   &   0.125  &   2.31 \\
   ZTF J134037.03+733037.0 \, & 205.154323   &  73.51027  &   0.266 \, &   0.1793  \, &   2.006   &   0.1    &   2.24 \\
   ZTF J202921.75+115328.4 \, & 307.340661   &  11.89123  &   0.571 \, &   0.2005  \, &   1.523   &   0.241  &   2.54 \\
   ZTF J223934.51+200033.8 \, & 339.893820   &  20.00940  &   0.308 \, &   0.1910  \, &   2.383   &   0.175  &   2.78 \\
   ZTF J210046.30+420324.7 \, & 315.192950   &  42.05683  &   0.276 \, &   0.2040  \, &   1.735   &   0.166  &   2.61 \\
   ZTF J055824.63+252036.2 \, & 89.6026576   &  25.34337  &   0.394 \, &   0.1530  \, &   1.766   &   0.188  &   2.55 \\
  CSS J081001.0+324429     \, & 122.504424   &  32.74127  &   0.35  \, &   0.1865  \, &   1.964   &   0.148  &   2.29 \\
   ZTF J215438.54+655322.6 \, & 328.660633   &  65.88961  &   0.273 \, &   0.1868  \, &   2.482   &   0.080  &   2.63 \\
   ZTF J165835.01+225905.7 \, & 254.645883   &  22.98490  &   0.317 \, &   0.1761  \, &   2.178   &   0.112  &   2.75 \\
   ZTF J135005.20+501920.7 \, & 207.521691   &  50.32240  &   0.156 \, &   0.2048  \, &   3.643   &   0.052  &   2.39 \\
   ZTF J204646.04+233929.0 \, & 311.691841   &  23.65805  &   0.546 \, &   0.1840  \, &   3.382   &   0.076  &   2.26 \\    
   \hline
  \end{tabular}
  \label{lowmass1}
  \end{center}
\end{table} 

The absolute magnitudes and other results for these systems are presented in Table\,\ref{lowmass2}. In order to calculate the absolute magnitude errors, the error propagation of both the magnitudes and parallaxes had to be taken into account. The errors in the distance moduli caused by the parallax uncertainties $\sigma_{p}$ were calculated and applied via:

\begin{eqnarray}
    \sigma_{m-M}=\sigma(-5\lg(p'')-5)\approx\frac{5}{\ln(10)}\frac{\sigma_{p}}{p}                           \\
    \sigma_{M_{G}}^{2}=\sigma_{m-M}^{2}+\sigma_{G}^{2} \\
    \sigma_{M_{J}}^{2}=\sigma_{m-M}^{2}+\sigma_{J}^{2}
\end{eqnarray}

The total galactic line-of-sight extinctions $A_{r}$ and $A_{J}$ were manually checked with the NED extinction calculator\footnote{\url{https://ned.ipac.caltech.edu/extinction\_calculator}}. The SDSS $r$ filter has a central wavelength close to the Gaia $G$ filter ($\approx600$\,nm). However, all selected objects are red, so they emit predominantly in the red region of the wide $G$-band, where the extinction is lower than in the center. So we can safely assume that the extracted $r$-band extinction values are upper limits, i.e. $A_{G}<A_{r}$.

As many of these objects are faint (18-20\,mag), there is a considerable risk of incorrect automatic classification of their variability type. Therefore each individual lightcurve was visually inspected. Comments of the variability types based on these manual inspections are also included in Table\,\ref{lowmass2}. The AAVSO VSX variability type designations are used for this purpose (EA = Algol (possibly with strong ellipsoidal effects), EB = $\beta$\,Lyr, EW = W UMa, ELL = rotating ellipsoidal variable). It appears that the VSX classification is false in at least 9 of the 46 cases and, possibly, in more than half of them.

The upper limits for the mass of the primary component, presented in the final column, are calculated as the lesser of the two main sequence masses, corresponding to $M_{Gmax}$ and $M_{Jmax}$, where:
\begin{eqnarray}
    M_{Gmax}=M_{G}-2\sigma_{M_{G}}-A-A_{r}                           \\
    M_{Jmax}=M_{J}-2\sigma_{M_{J}}-A-A_{J}                           
\end{eqnarray}

The absolute magnitudes and masses are compared to intepolated values from Table\,\ref{Mamajek}.

These are almost worst-case-scenario absolute magnitude values -- the amplitudes $A$ are subtracted in case the $G$ and $J$-band estimates are made near a minimum, and maximum extinctions are considered, although most systems are nearby (Fig.\,\ref{fig_period}). Since coefficients of 2 are used for taking the absolute magnitude errors into account, these upper limits are set with a $\sim95\%$ confidence, i.e. a roughly normal probability distribution is assumed for the absolute magnitudes and for the mass of the primary. These mass limits are calculated only for the systems, whose lightcurves correspond to a W UMa-type, and are valid only in case the systems actually are contact binaries. A total of 26 systems in Table\,\ref{lowmass2} are expected to have a primary component less massive than $0.5M_{\sun}$.

\begin{table}[!htb]
\scriptsize
  \begin{center}
  \caption{The list of selected M-type contact binary candidates with the lowest $J$-band luminosity: calculated absolute magnitudes (uncorrected for extinction), maximum extinction values, lightcurve classification and upper limits for the mass of the primary.}
  \begin{tabular}{lcccccccc}
  \hline
   Name                     \, & $M_G$ \, & $\sigma_{M_{G}}$  \, &  $M_J$ \, &  $\sigma_{M_{J}}$  \, & $A_{r}$   \, &  $A_{J}$  \, &  lightcurve \, & $M_{1 max}$[\msol] \\           
  \hline                                                                                                                              
   ZTF J024005.10+513845.5  \, &  12.22 &   0.43  \, &  8.977  &   0.43  \, &    0.71  &  0.22  \, &  noisy            &   $<0.35$  \\ 
   WTS 19b-3-06008          \, &  11.99 &   0.29  \, &  8.620  &   0.29  \, &    0.40  &  0.12  \, &  EW               &   $<0.32$  \\ 
  CSS J093044.8+264957      \, &  10.80 &   0.38  \, &  8.410  &   0.39  \, &    0.05  &  0.02  \, &  EW	       &   $<0.37$  \\ 
   ZTF J150926.26+583628.1  \, &  11.12 &   0.01  \, &  8.110  &   0.03  \, &    0.02  &  0.01  \, &  EA/EB            &     -      \\ 
   ZTF J001012.42+452925.1  \, &  10.78 &   0.05  \, &  8.092  &   0.07  \, &    0.18  &  0.06  \, &  noisy/sparse     &   $<0.33$  \\ 
   ZTF J213409.46+335033.8  \, &  10.82 &   0.16  \, &  7.748  &   0.16  \, &    0.40  &  0.13  \, &  EB               &     -      \\ 
   ZTF J191844.32+532028.2  \, &  10.37 &   0.06  \, &  7.657  &   0.07  \, &    0.22  &  0.07  \, &  EA/EB            &     -      \\ 
   ZTF J233629.58+505049.5  \, &  10.77 &   0.02  \, &  7.657  &   0.03  \, &    0.44  &  0.14  \, &  noisy/sparse     &   $<0.36$  \\ 
   ZTF J204919.36+080812.0  \, &  10.31 &   0.20  \, &  7.629  &   0.21  \, &    0.17  &  0.05  \, &  EB/EW            &   $<0.40$  \\ 
   ZTF J175927.14+654540.4  \, &  10.69 &   0.07  \, &  7.626  &   0.08  \, &    0.11  &  0.04  \, &  EB               &     -      \\ 
  CSS J234019.3+122438      \, &  10.19 &   0.02  \, &  7.586  &   0.03  \, &    0.12  &  0.04  \, &  EW/ELL           &   $<0.39$  \\ 
   ZTF J174317.23+401313.1  \, &  10.80 &   0.20  \, &  7.575  &   0.22  \, &    0.08  &  0.03  \, &  EA/EB            &     -      \\ 
   ZTF J145410.58+314216.2  \, &  10.42 &   0.01  \, &  7.568  &   0.02  \, &    0.04  &  0.01  \, &  EA/EB            &     -      \\ 
   ZTF J193121.28+293951.2  \, &  10.33 &   0.17  \, &  7.516  &   0.19  \, &    1.32  &  0.41  \, &  EW               &   $<0.51$  \\ 
   ZTF J051637.22+380939.8  \, &  9.382 &   0.37  \, &  7.491  &   0.40  \, &    1.83  &  0.57  \, &  EW               &   $<0.58$  \\ 
   ZTF J220421.03+163655.8  \, &  10.40 &   0.08  \, &  7.476  &   0.08  \, &    0.12  &  0.04  \, &  EW               &   $<0.39$  \\ 
   ZTF J174214.33+642525.0  \, &  10.38 &   0.03  \, &  7.450  &   0.04  \, &    0.08  &  0.03  \, &  EA/EB            &     -      \\ 
   ZTF J182843.57+372540.6  \, &  9.637 &   0.28  \, &  7.426  &   0.31  \, &    0.11  &  0.04  \, &  EB/EW            &   $<0.50$  \\ 
   ZTF J011509.92+130340.9  \, &  9.743 &   0.36  \, &  7.388  &   0.39  \, &    0.08  &  0.03  \, &  EW               &   $<0.54$  \\ 
   ZTF J015312.72+444309.8  \, &  10.36 &   0.04  \, &  7.372  &   0.05  \, &    0.17  &  0.05  \, &  EW               &   $<0.40$  \\ 
   ZTF J003651.36+503617.7  \, &  10.11 &   0.19  \, &  7.351  &   0.20  \, &    0.38  &  0.12  \, &  noisy/sparse     &   $<0.48$  \\ 
   ZTF J175432.47+373447.6  \, &  10.39 &   0.01  \, &  7.344  &   0.03  \, &    0.09  &  0.03  \, &  EW               &   $<0.37$  \\ 
   ZTF J053128.93+512221.6  \, &  9.990 &   0.37  \, &  7.308  &   0.38  \, &    0.84  &  0.26  \, &  EW/ELL           &   $<0.53$  \\ 
  CSS J082958.5+355923      \, &  10.35 &   0.05  \, &  7.287  &   0.05  \, &    0.10  &  0.03  \, &  EW/ELL           &   $<0.38$  \\ 
   ZTF J011434.64+622754.1  \, &  10.17 &   0.03  \, &  7.275  &   0.05  \, &    3.54  &  1.10  \, &  EW               &   $<0.58$  \\ 
   ZTF J221742.28+485418.4  \, &  10.08 &   0.06  \, &  7.237  &   0.07  \, &    0.67  &  0.21  \, &  EW/ELL           &   $<0.45$  \\ 
   ZTF J183104.30+712050.5  \, &  10.06 &   0.04  \, &  7.226  &   0.05  \, &    0.12  &  0.04  \, &  noisy            &   $<0.40$  \\ 
  CSS J111647.8+294602      \, &  10.02 &   0.04  \, &  7.219  &   0.05  \, &    0.04  &  0.01  \, &  EW/ELL           &   $<0.42$  \\ 
   ZTF J001725.54+513930.3  \, &  10.02 &   0.16  \, &  7.213  &   0.17  \, &    0.38  &  0.12  \, &  EW/ELL           &   $<0.48$  \\ 
   2MASS J08115860+3119595  \, &  9.579 &   0.02  \, &  7.186  &   0.03  \, &    0.09  &  0.03  \, &  EB/EW            &   $<0.44$  \\ 
   ZTF J112409.81+314128.5  \, &  9.969 &   0.18  \, &  7.158  &   0.18  \, &    0.04  &  0.01  \, &  EW               &   $<0.45$  \\ 
  CSS J224134.8+253648      \, &  10.37 &   0.02  \, &  7.154  &   0.03  \, &    0.11  &  0.03  \, &  EW/ELL           &   $<0.38$  \\ 
  CSS J030238.1-065034      \, &  10.05 &   0.05  \, &  7.123  &   0.06  \, &    0.18  &  0.05  \, &  EW               &   $<0.42$  \\ 
  CSS J001242.4+130809      \, &  10.07 &   0.01  \, &  7.118  &   0.02  \, &    0.19  &  0.06  \, &  EB/EW            &   $<0.41$  \\ 
   ZTF J184257.77-014740.2  \, &  10.05 &   0.12  \, &  7.117  &   0.14  \, &    7.58  &  2.35  \, &  EW               &   $<0.70$  \\ 
   ZTF J191410.07+322037.4  \, &  9.865 &   0.14  \, &  7.085  &   0.16  \, &    0.40  &  0.13  \, &  noisy            &   $<0.48$  \\ 
   ZTF J134037.03+733037.0  \, &  9.515 &   0.10  \, &  7.082  &   0.12  \, &    0.04  &  0.01  \, &  EB/EW            &   $<0.47$  \\ 
   ZTF J202921.75+115328.4  \, &  9.933 &   0.34  \, &  7.080  &   0.35  \, &    0.21  &  0.07  \, &  EB               &     -      \\ 
   ZTF J223934.51+200033.8  \, &  10.11 &   0.16  \, &  7.070  &   0.16  \, &    0.08  &  0.03  \, &  EW               &   $<0.44$  \\ 
   ZTF J210046.30+420324.7  \, &  10.04 &   0.20  \, &  7.068  &   0.22  \, &    3.10  &  0.96  \, &  EW               &   $<0.67$  \\ 
   ZTF J055824.63+252036.2  \, &  9.781 &   0.23  \, &  7.067  &   0.24  \, &    2.61  &  0.81  \, &  EW               &   $<0.67$  \\ 
  CSS J081001.0+324429      \, &  9.592 &   0.16  \, &  7.066  &   0.17  \, &    0.13  &  0.04  \, &  EW/ELL           &   $<0.52$  \\ 
   ZTF J215438.54+655322.6  \, &  9.857 &   0.07  \, &  7.064  &   0.08  \, &    1.36  &  0.42  \, &  EA/EB            &      -     \\ 
   ZTF J165835.01+225905.7  \, &  9.996 &   0.11  \, &  7.051  &   0.12  \, &    0.17  &  0.05  \, &  EB/EW            &   $<0.45$  \\ 
   ZTF J135005.20+501920.7  \, &  9.664 &   0.03  \, &  7.019  &   0.04  \, &    0.03  &  0.01  \, &  EB/EW            &   $<0.45$  \\ 
   ZTF J204646.04+233929.0  \, &  9.756 &   0.04  \, &  7.010  &   0.05  \, &    0.36  &  0.11  \, &  EW               &   $<0.51$  \\ 
   \hline
  \end{tabular}
  \label{lowmass2}
  \end{center}
\end{table}

\section*{Summary}
The absolute magnitudes of main sequence M-dwarfs are very sensitive of their masses. The primary components in contact binary systems are expected to have a luminosity not lower than that of a main sequence star with the same mass. Thus we can set an upper limit of the mass of the primary only from the absolute magnitude of the system. This is especially useful in the search for M-type contact binaries since most of these objects are faint and difficult to observe spectroscopically. This paper aims to present such an approach for selecting promising targets for more detailed exploration. By combining Gaia parallaxes with Gaia and 2MASS photometry we selected a total of 218 systems below the $0.22$\,d period cutoff as potential M-type contact binaries. A smaller subselection of 46 of these systems was manually inspected -- the classification of their lightcurves was checked and the masses of the primaries were limited. Future targeted observations of these systems will show their exact location on the low-mass end of the distribution of contact binary systems and how they got there.

\section*{Acknowledgements}
This work has made use of data from the European Space Agency (ESA) mission
{\it Gaia} \\ (\url{https://www.cosmos.esa.int/gaia}), processed by the {\it Gaia}
Data Processing and Analysis Consortium (DPAC\footnote{\url{https://www.cosmos.esa.int/web/gaia/dpac/consortium}}). Funding for the DPAC
has been provided by national institutions, in particular the institutions
participating in the {\it Gaia} Multilateral Agreement. This work was supported by the Bulgarian Ministry of Education and Science under the National Research Programme "Young scientists and postdoctoral students" approved by DCM\,\#577/17.08.2018.


\end{document}